\documentclass[preprint,amsmath,amssymb,superscriptaddress,showpacs]{revtex4}
\usepackage{graphicx}
\begin{document}
\baselineskip=0.8cm
\renewcommand{\thesection}{\arabic{section}}
\renewcommand{\thesubsection}{\arabic{section}.\arabic{subsection}}
\renewcommand{\thefigure}{\arabic{figure}}
\baselineskip=0.7cm
\title{Ground-state properties of the one dimensional electron liquid}
\author{R. Asgari\\
\it{Institute for Studies in Theoretical Physics and
Mathematics, Tehran 19395-5531, Iran}}
\begin{abstract}
We present a theory of the pair distribution function $g(z)$ and
many-body effective electron-electron interaction for one
dimensional (1D) electron liquid. Our approach involves the solution
of a zero-energy scattering Schr\"odinger equation for $\sqrt{g(z)}$
where we implemented the Fermi hypernetted-chain approximation
including the elementary diagrams corrections. We present numerical
results for $g(z)$ and the static structure factor $S(k)$ and obtain
good agreement with data from diffusion Monte Carlo studies of the
1D system. We calculate the correlation energy and charge excitation
spectrum over an extensive range of electron density. Furthermore,
we obtain the static correlations in good qualitative agreement with
those calculated for the Luttinger liquid model with long-range
interactions.
\end{abstract}
 \maketitle

\section{Introduction}
\label{sec1}

Electron-electron interaction is known to produce the most
pronounced effects in one dimensional electron liquid (1D EL)
systems, where a strongly correlated state appears~\cite{voit}. The
effects of interactions in higher than one dimensional systems has
been masterfully explained by Landau's Fermi liquid
theory~\cite{landau} dealing with low-lying excitations in a system.
Landau called such single-particle excitations quasiparticles and
postulated a one-to-one correspondence between the quasiparticles
and the excited states of a non-interacting Fermi gas. However, for
the 1D EL any individual excitation becomes a collective one. This
collectivization of excitations is obviously a major difference
between the 1D and higher dimensions electron liquids. It clearly
invalidates any possibility to have a Fermi liquid theory in 1D EL.
One of the most interesting characteristics of the system is the
spin-charge separation seen in a series of remarkable
experiments~\cite{auslaender}. Furthermore, according to the theorem
proved by Lieb and Mattis~\cite{lieb}, the ground state of a 1D EL
is paramagnetic. The recent quantum Monte Carlo
calculations~\cite{malatesta} consonant with the Lieb and Mattis
theorem predicts no Bloch instability~\cite{bloch} for the 1D EL.

Many of the electron-electron interaction effects become
increasingly important as carrier density and dimensionality are
reduced and the homogenous electron liquid provides a primitive
model for their study. The crucial role in the theory is played by
the particle pair distribution function $g(z)$. In our previous
works we presented a theory of the pair distribution function and
other ground-state properties of the 3D and 2D electron
liquid~\cite{davoudi,asgari}. The theory was based on a Fermi
hypernetted-chain approximation (FHNC) which represents a direct
generalization to the Fermi systems of the well known
hypernetted-chain Euler-Lagrange (HNC/EL) approximation. One of the
important feature of theory based on HNC/EL or FHNC is that the
theory sums not only all ring and ladder diagrams exactly, but also
mixed diagrams in a local approximation~\cite{jackson}. It means the
FHNC theory is not a Fermi liquid theory and therefore we able to
apply it for studying many-body effects on 1D EL. As we shall see,
much higher sophistication is needed to attain a quantitatively
useful theory in 1D EL. We shall have to dwell on terms beyond the
FHNC/0, which includes elementary diagrams and the three-body
Jastrow-Feenberg correlations. These effects have been studied
theoretically in boson fluids for the 1D system using the HNC/EL
formalism by Krotscheck {\it et al}~\cite{krotscheck}.

The main purpose of the present work is the formulation of a
practicable theory for charge correlations in the paramagnetic 1D EL
and comparison of its numerical predictions with the available
diffusion Monte Carlo data~\cite{casula}. The theory is based on
variational FHNC~\cite{lantto80} method made by simplifying the
standard FHNC method to have only one Euler-Lagrange Schr\"odinger
equation. The computational time typically needed to obtain $g(z)$
at each value of density is a few minutes on a PC with a Pentium
IV/3.4 GHz processor. There are many theoretical works addressing
the ground-state properties of the quasi- and
strictly~\cite{gold,gold1} 1D EL but an accurate charge correlations
is still missing.

In this work, we numerically study the many-body effects for 1D EL
by calculating the pair distribution function and electron-electron
effective interactions.

\section{Theory}
\label{sec2}
 We consider a 1D EL as a model for a system of
electronic carriers with band mass $m$ in a semiconductor
heterostructure with dielectric constant $\epsilon$. The resulting
effective 1D potential is readily shown~\cite{friesen} to be
$v(z)=(e^2/\epsilon)(\sqrt{2}/2b)~exp[z^2/4b^2]~erfc[\left|z\right|/2b]$
with Fourier transform $v(k)=(e^2/\epsilon)~n~
exp[b^2k^2]~E_1[b^2k^2]$ where the exponential integral function,
$E_1(x)$ is defined as $\int_x^{\infty} e^{-u} du/u$. The Fourier
transform is defined according to the general expression $FT[F(r)]=
n\int d{\bf r} F(r) \exp{(i{\bf k}\cdot {\bf r})}$. Here $n$ is the
total average density. The above form of the bare potential exhibits
the typical 1D behavior, $v(k)\approx ln(k b)$ as $k\rightarrow 0$,
and the 3D behavior, $v(k)\approx 1/(k b)^2$ as $k\rightarrow
\infty$.
 At zero temperature there are only two relevant parameters for a
disorder-free, homogeneous 1D EL in the absence of quantizing
magnetic fields, namely the quantum wire width $b$ and the usual
Wigner-Seitz density parameter $r_s=(2 n a^*_B)^{-1}$, where
$a^*_B=\hbar^2 \epsilon/(m e^2)$ is the Bohr radius in the medium of
interest.
 We assume that only one subband
is occupied and neglect any contribution from higher subbands. This
approximation leads to $r_s > \pi b/4$.

With the zero of energy taken at the chemical potential, the
formally exact Euler-Lagrange equation for the spin summed pair
distribution function $g(z)$ reads~\cite{lantto, kallio}
\begin{equation}\label{e5}
\left[-\frac{\hbar^2}{m}~\frac{d^2}{dz^2}+v(z)+
v_B(z)+
v_F(z)\right]
\sqrt{g(z)}=0\,.
\end{equation}
Here, $v(z)$ is the 1D EL potential and the Bose-like potential
$v_B(z)$ contains the effects of correlations and by itself would
determine $g(z)$ in a Bose fluid. The details of the Bose-like
potential have been discussed by many authors, for instance see Ref.
[\onlinecite{report}]. We can write it as
\begin{equation}\label{vb}
v_B(z)=w_{ind}(z)+\Delta v_{ele}(z)~,
\end{equation}
where the "induced interaction" is~\cite{apaja}
\begin{equation}\label{ind}
w_{ind}(k)=-\frac{\hbar^2k^2}{4m}\left[S(k)-1\right]-\tilde{v}_{ph}(k)~.
\end{equation}
A momentum space formulation of the Euler equations equivalent
to Eq.~(\ref{e5}) can be written in terms of the structure factor $S(k)$,
\begin{equation}\label{sq}
S(k)=\frac{1}{\sqrt{1+\frac{4m}{\hbar^2k^2}\tilde{v}_{ph}(k)}}~,
\end{equation}
and the FHNC-Euler-Lagrange theory supplements Eq.~(\ref{sq}) with a
microscopic theory of the particle-hole interaction~\cite{report}
\begin{equation}\label{vph}
\tilde{v}_{ph}(z)=g(z)[v(z)+\Delta v_{ele}(z)+v_F(z)]+\frac{\hbar^2}{m}\left| \frac{d}{dz}\sqrt{g(z)}\right|^2+[g(z)-1]w_{ind}(z)~.
\end{equation}
From the theoretical point of view it is important to remark that in
FHNC-type calculations at strong coupling the Bose-like $v_{B}(q)$
interactions should be corrected by the addition of three-body
correlations and elementary-diagrams (or "bridge functions")
contributions~\cite{lantto,zab,apaja,report}. The $\Delta
v_{ele}(z)$ is a term arising from elementary diagrams and triplet
correlations. We only consider the fourth-order elementary diagrams
and triplet correlations, $\Delta
v_{ele}(k)=w_B^{E_4}(k)+w_B^{(3)}(k)$, which hereafter, we call it
as FHNC$/4+$triplet acronym. Note that the approach is reduced to
the Fermi hypernetted-chain approximation, FHNC$/0$ when the
corrections of the Bose-like interaction are omitted, $\Delta
v_{ele}(k)=0$. The contribution from the low-order elementary
diagrams to the effective Bose potential is
\begin{equation}
w_B^{E_4}(k)=-\frac{\hbar^2}{4mn}\left\{\frac{k^2}{2\pi}\varepsilon_4(k)+\int\frac{dq}{(2\pi)}q^2[S(q)-1]\frac{\delta\varepsilon_4(q)}{\delta S(k)}\right\},
\end{equation}
where $\varepsilon_4(q)$ is given by a twofold integral
(in $1D$) over momentum space.
\begin{equation}
\varepsilon_4(q)= \frac{1}{2\,n^4}\int \frac{dp\,dq^{'}}{(2\pi)^2}[S(p)-1]
[S(q^{'})-1][S(p+q^{'})-1][S(p+q)-1]~[S(p+q^{'}+q)-1]~,
\end{equation}
and the contribution of three-body correlations is given by the integral,
\begin{eqnarray}
w_B^{(3)}(k)=\frac{1}{8n\pi}\int d{q}S(p)S(q)
u_3({q},{p},{k})
\left\{v({q},{p},{k})
+\left[\varepsilon(p)+\varepsilon(q)\right]u_3({q},{p},{k})\right\}
\end{eqnarray}
where ${\bf p}=-({q}+{k})$, $\varepsilon(k)=\hbar^2k^2/[2mS(k)]$
and, with the definition $X(k)=1-S^{-1}(k)$,
we have $v({q},{p},{k})=(\hbar^2/m) \left[{k}{p}X(p)+{k}{q}X(q)+{p}{q}X(p)X(q)\right]$ and
\begin{equation}
u_3({q},{p},{k})=-\frac{(\hbar^2/2m)\left[{k}{p}X(p)X(k)+{p}{q}X(p)X(q)+{k}{q}X(q)X(k) \right]}{\varepsilon(k)+\varepsilon(p)+\varepsilon(q)}~.
\end{equation}
With regard to the Fermi term $v_F(z)$ in Eq. (\ref{e5}), we adopt
the same criteria in determining its form as in
Ref.~[\onlinecite{kallio}]. An important requirement is that
Eq.~(\ref{e5}) should give the exact fermion distribution function
when one goes to the weak-coupling limit $r_s\rightarrow 0$, where
in this limit $g(z)$ becomes the Hartree-Fock pair distribution
function, $g^{\rm \scriptscriptstyle HF}(z)$. The Fermi term in the
scattering potential is then determined by the Hartree-Fock
structure factor $S^{\rm \scriptscriptstyle HF}(k)$,
\begin{equation}\label{ef}
v_F(k)=\frac{\hbar^2}{m}{\rm FT}\left[ \frac{d^2\sqrt{g^{\rm \scriptscriptstyle HF}(z)}/dz^2}{\sqrt{g^{\rm \scriptscriptstyle HF}(z)}}\right]+
\frac{\hbar^2k^2}{4m}\left[\frac{S^{\rm \scriptscriptstyle HF}(k)-1}{S^{ \rm \scriptscriptstyle HF}(k)}\right]^2
[2S^{\rm \scriptscriptstyle HF}(k)+1]-\Delta v_{ele}(k)|_{S(k)=S^{\rm \scriptscriptstyle HF}(k)}.
\end{equation}
The Hartree-Fock static structure function is given by $S^{\rm
\scriptscriptstyle HF}(k)=(k/k_F)\theta(2k_F-k)+\theta(k- 2k_F)$
where $\theta$ is the Heaviside step function, $k_F=\pi/4 r_s a^*_B$
is the Fermi wave vector and $g^{\rm \scriptscriptstyle HF}(z)$ is
given by
\begin{equation}
g^{\rm \scriptscriptstyle HF}(z)=
1-\frac{sin^2(k_F z)}{2\,(k_F z)^2}~.
\end{equation}
 The rational behind Eq.~(\ref{ef}), is as in Ref.~[\onlinecite{kallio}],
 (i) the first term on the RHS ensures that the Hartree-Fock limit
 is correctly embodied into the theory, (ii) the second and third
 terms ensure that the Bose-like scattering potential is
 suppressed for  electrons at weak coupling~\cite{davoudi}.
 We introduce at this
point the relation between $g(z)$ and $S(k)$:
\begin{equation}
g(z)=1+\frac{2}{n}\,\int \,cos(k\,z)
[S(k)-1]\,dk~.
\end{equation}

\subsection{Sum rules and limiting behaviors}

In this section we show that the pair distribution function obtained
from the theory presented in Sec. II satisfy three exact properties.
These are the behavior of the static structure factor at small
momentum, the charge neutrality and the Kimball's cusp condition.

The asymptotic behavior of the effective potential $v_{ph}(z)$ is
first obtained from Eq.~(\ref{vph}). A careful analysis of this
equation shows that $v_{ph}(z)\rightarrow v(z)+v_F(z)$ for
$z\rightarrow\infty$, and therefore in Fourier transformation we
have $v_{ph}(k)\rightarrow v(k)+\hbar^2 k_F^2/m$ for $k\rightarrow
0$, with $v(k)$ as the Fourier transform of bare potentail the 1D
EL. The corresponding asymptotic behavior of the structure factor is
obtained from Eq.~(\ref{sq}), which we have
\begin{equation}
S(k)\rightarrow \frac{\hbar k}{\sqrt{4 m v(k)}}~,
\end{equation}
for $k\rightarrow 0$. The analytical behavior of correlations for
the 1D EL with long range potential interaction has been obtained by
Schulz~\cite{schulz} using the bosonization techniques applied to an
effective 1D hamiltonian with linearized kinetic energy expression.
According to that model, $S(k)$ behaves like $\propto
k/\sqrt{\left|ln(k)\right|}$. It is clear that our scheme gives
\begin{equation}\label{lims}
lim_{k\rightarrow 0}S(k)=k\sqrt{\frac{r_s}{4 \left|ln(k)\right|}}~.
\end{equation}
This result agrees with the bosonization findings in 1D.
The form of Eq. ~(\ref{lims}) immediately ensures that the
charge neutrality condition is satisfied. The charge neutrality condition,
\begin{equation}
n\int [g(z)-1]dz=-1~,
\end{equation}
is equivalent to $S(k\rightarrow 0)=0$, a result obtained from
Eq.~(\ref{lims}). We have numerically evidence that the charge
neutrality condition is fulfiled for our total effective potential
in Eq.~(\ref{e5}). On the other hand, in the small$-r$ limit the
bare repulsion potential dominates over the induced plus Fermi
potentials. Therefore, it follows from the differential
Eq.~(\ref{e5}) that
\begin{equation}
g(z\rightarrow 0)= g(0)\left(1+\frac{r_s}{12~ b}\, z\right)~.
\end{equation}
This is a similar result of the Kimball's cusp
condition~\cite{kimball} for 3D electron gas. Linear small$-z$
behavior together with general properties of Fourier transformations
implies that at large $k$ the static structure function $S(k)-1$
approaches zero asymptotically in proportion to $-1/k^4$.

\subsection{Charge excitations in long-wavelength limit}

From the knowledge of static structure factor we can find the behavior of
the charge excitations in the long-wavelength limit. Generally, one writes the
exact response function as
\begin{equation}
\chi(k,\omega)=\frac{\chi_0(k,\omega)}{1-v(k)(1-G(k,\omega))\chi_0(k,\omega)}~,
\end{equation}
where $\chi_0(k,\omega)$ is the response function\cite{gv_book} of
the noninteracting 1D EL and the $G(k,\omega)$ is the charge-charge
dynamic local-field factor. In this work, we approximate the dynamic
local-field factor by its $\omega-$independent, G(k). The charge
excitation is determined by calculating the pole of response
function. The  dispersion relation of charge excitations is given by
\begin{equation}\label{wq}
\omega(k)=\left[\frac{A~\beta^2_{+}-\beta^2_{-}}{(A-1)}\right]^{1/2}~,
\end{equation}
where $A=exp(\pi\hbar^2 k/m v(k)(1-G(k))$ and $\beta_{\pm}=\hbar(k^2/2\pm k_F k)/m$.

 In the limit $k\rightarrow 0$, the effective potential in
 momentum space, $v(k)(1-G(k))$ reduce to Fourier transformation
 of bare potential and $\omega(k)$ turns out to be
\begin{equation}\label{wk}
\hbar \omega(k)=\frac{e^2}{a^*_B}~ k~ \sqrt{\frac{\left|ln(k)\right|}{r_s}}~.
\end{equation}
This behavior of the $\omega(k)$ is independent of the statistics
due to the exchange potential, $v_F(r)$ is of shorter range than
$v_B(r)$ and essentially the low-momentum limit of $\omega(k)$
remains unaltered. The expression can also be written as:
\begin{equation}
\omega(k\rightarrow 0)=\frac{\hbar k^2/2\,m}{S(k)}\nonumber~.
\end{equation}
On the other hand, the charge oscillation eigenmodes provided by the
bosonization model hamiltonian of the Coulomb interaction
liquid~\cite{schulz,wang} is given by
\begin{equation}
\omega_+(k)=\frac{\hbar k k_F}{m}\,\sqrt{(1+\bar{g_1})(1-\bar{g_1}+2mv(k)/\pi k_F)}~,
\end{equation}
where $\bar{g_1}=g_1 k_F/2m\pi$ and $g_1$ is a nonsingular
interaction matrix element at $2k_F$. For small $k$ excitations, we have
\begin{equation}
\omega_+(k)=\frac{e^2}{a^*_B}\sqrt{1+\bar{g_1}}~k \,\sqrt{\frac{\left|ln(k)\right|}{r_s }}
\end{equation}
which is very similar to our finding in Eq.~(\ref{wk}).

\subsection{Fluctuation-dissipation theorem}

In order to calculate the charge excitation spectrum from
Eq.~(\ref{wq}), the local-field factor is needed. We construct the
local-field factor from the fluctuation-dissipation theorem. This
theorem, which is of paramount importance for systems in
equilibrium, relates the dynamic susceptibility defined above to the
static structure factor
\begin{equation}\label{lff}
S(k)=-\frac{1}{\pi}\,\int^{\infty}_0 d\omega\,
\Im[\chi(k,\omega)]~,
\end{equation}
As $\chi(k,\omega)$ depend on $G(k)$ the above integral expression
allows one to determine the local-field factor once the static
structure factor is calculated. The same approach of obtaining the
local-field factor has previously been employed by
Iwamoto~\cite{iwamoto} and Dharma-wardana and
Perrot~\cite{dharma-wardana}. We note that the use of
fluctuation-dissipation theorem to extract static local-field factor
is approximate in nature as it neglects the frequency dependence of
$G(k,\omega)$ from the outset. However, the apparent success of our
previous work in 2D electron gas context~\cite{asgari_2} also
encourages us to use it in the present work.

\section{Numerical Results}
 We turn to a presentation of our numerical results, which
 are obtained by solving Eq.~(\ref{e5}) with the following
 self-consistency cycle. We start with the trial choice
 $g(z)=g^{\rm \scriptscriptstyle HF}(z)$ and $v_B(z)=0$,
 and find the effective potentials $\tilde{v}_{ph}(z)$ by
 means of Eq.~(\ref{vph}) and hence the structure factor
 $S(k)$ via Eq.~\ref{sq}. At this point we can calculate
 new values for $g(z)$ and for $v_{B}(z)$ by taking Fourier
 transforms and using Eqs.~(\ref{vb}) and (\ref{ind}).
 This procedure is repeated until self-consistency is achieved.
 We have calculated in this way the pair distribution function
 of a 1D EL for various $r_s$ and $b$ values. The main results
 of our work are shown in Figures 1-6.

Our numerical results of the pair distribution function of the 1D EL
is compared with the DMC data of Ref.~[\onlinecite{casula}] in
Figure 1. The quantum wire width is $b=0.1 a^*_B$ and the system is
highly correlated. We have clearly achieved good qualitatively
agreement with the DMC data. It is obvious from the Fig.~1,
FHNC$/4+$triplet results are in excellent agreement with DMC data up
to $z k_F <1.3$ and less agreement in larger $z$-values to produce
the oscillation behavior. The strong oscillation behavior in DMC
data corresponds to the periodicity of the quasi-Wigner crystal. It
is apparent from the figure that
 FHNC$/4+$triplet approach which includes further corrections
 than FHNC/0, modifies the results and show the first peak and
 its oscillatory behavior, however the magnitude of peak is
 less than the one predicted by DMC simulation. Our $g(z)$
 has more structure within FHNC$/4+$triplet than the $g(z)$
 obtained within FHNC/0. Moreover, in both approaches the
 short-range behavior of $g(z)$ give correct shape in base
 on Kimball's cusp condition and they have different values at contact.

In Figure 2, we show our results for $S(k)$ in the paramagnetic 1D
EL at $r_s=1$ and $2$ within both FHNC$/0$ and FHNC$/4+$triplet
approximations. It is obvious to see the improvements brought above
by the use of FHNC$/4+$triplet over  FHNC/0. Apart from the strong
peak structure at $q=4 k_F$ which appears in DMC simulation, the
FHNC$/4+$triplet results are in good agreement with the DMC data of
Casula {\it et al}~\cite{casula}. The strong peak has been related
to a quasi-order of the electrons~\cite{schulz,casula}. Note that
there is no true long-range order in 1D system. We find that as the
density is reduced the correlation effects become stronger and
$S(k)$ starts to develop a broad peak around $q=4.5 k_F$. The
FHNC$/0$ does not show any peak as the density decreases. Theory
gives the correct behavior of the long-wavelength limit of $S(k)$
and the results coincide with the DMC data.

There are several noteworthy points based on the results shown in
Figs.~1 and 2. (i) Despite that FHNC gives very good results for the
pair distribution function and static structure factor in high
dimensional electron liquids~\cite{davoudi,asgari,lantto80,zab}, it
can give qualitative results for 1D EL where the system is highly
correlated. (ii) The approach could not produce a strong peak for
static structure factor at $q=4k_F$ which is related to the slow
decay of the $4k_F$ components of the charge density-density
correlation function.

In order to emphasize the effect of correlations at strong coupling
regime, we display in Fig.~3 the static structure factor obtained
within both FHNC$/0$ and FHNC$/4+$triplet approximations. The strong
coupling arises with decreasing either the density of electrons or
the quantum wire width parameter $b$. $S(k)$ move towards right in
intermediate of momenta value and starts to develop a broad peak
around $q=4.5 k_F$ within the FHNC$/4+$triplet approach. In the
inset of the Fig.~3a, the behavior of the long-wavelength limit of
$S(k)$ both in FHNC/0 and FHNC$/4+$triplet approximation are shown
in comparison with sum-rule given by Eq. (\ref{lims}). As the
sum-rule of the long-wavelength limit of $S(k)$ is proven in the
based on our approach, we numerical evidence that this sum-rule is
fulfilled. It is apparent from Fig.~3b that by reducing the $b$
value, the peak moves towards right and the magnitude of the peak
becomes larger for the same $r_s$ value.

The corresponding pair distribution functions are shown in Fig.~4.
Again, we show the effect of correlations which are corrected by
including further correction functions to the Bose potential within
the FHNC$/4+$triplet approximation. The value of $g(0)$ decreases
with decreasing either the density $n$ or $b$ values, but its value
always remains positive, consistent with the probabilistic
definition. Moreover, In the inset of the Fig.~4a, the behavior of
the short range limit of $g(z)$ both in our approaches are shown in
comparison with cusp condition. Note that the $g(0)$ value at
contact is dependence the approach. We numerical evidence that the
cusp condition is fulfilled.

In Fig.~5 we report the effective electron-electron interaction
$v(z)+v_B(z)+v_F(z)$ as it emerges from our self-consistent
calculations on a 1D EL. The attractive part of the effective
interaction deepens with increased $r_s$ or decreased $b$, as is
physically expected. Moreover, its deepens in more larger in
FHNC$/4+$triplet than FHNC/0.

After reaching good agreement between the static structure factor
with data from diffusion Monte Carlo, we can then calculate the
correlation energy per particle $\varepsilon_c(r_s)$, the difference
between the exact ground-state energy and the sum of the
noninteracting kinetic energy and the exchange energy, of the 1D EL
as a function of $r_s$, using an integration over the coupling
strength according to the expression
\begin{equation}
\varepsilon_c(r_s)=\frac{1}{2}\int_{0}^{1}
\frac{d\lambda}{\lambda}~\int \frac{d k}{2
\pi}~v^{(\lambda)}_k~[S_{\lambda}(k)-S^{\rm \scriptscriptstyle
HF}(k)]~,
\end{equation}
(in Rydberg units). Here the static structure factor
$S_{\lambda}(k)$ calculated with interaction
$v^{(\lambda)}_k=\lambda (e^2/\epsilon)~ exp[b^2k^2]~E_1[b^2k^2]$.
We have calculated the correlation energy for $b=a^*_B$ and $2a^*_B$
over the range $0.1 \leq r_s \leq 4$. The results are reported in
Table I in comparison with DMC data from Casula {\it et
al}.~\cite{casula} The Table also include other theoretical results
obtained in the self-consistent dielectric theory of Singwi {\it et
al.}~\cite{stls} with an analytical form of the local-field factor
from Calmels and Gold.~\cite{gold}

It is seen from Table that the present theoretical approach yields
fairly better accurate values of the correlation energy respect to
the Calmels and Gold results~\cite{gold}.

Finally, in Fig.~6 we display the charge excitation spectrum
obtained within the FHNC$/4+$triplet approximation. The $\omega(k)$
behaviors like $k\sqrt{\left|lnk\right|}$ at long-wavelength limit
and reaches to the particle-hole continuum boundary which is given
by $\beta_+$ in the large $k$-limit. We use the local-field factor
which is calculated from the fluctuation-dissipation theorem by
Eq.~(\ref{lff}). The local-field factor should be useful to study
many physical characteristics using a suitably defined the
dielectric function.

\section{Summary}
In summary, we have presented in this work a theoretical study of
the pair distribution function, the effective electron-electron
interaction and the correlation energy of the 1D electron liquid
where the system is strongly correlated. Our approach yields
numerical results of good agreement in comparison with recent
diffusion Monte Carlo studies\cite{casula}. We showed that the
long-wavelength behavior of the static structure factor is in
agreement with the bosonization findings for the Coulomb Luttinger
liquid. The charge excitation spectrum is calculated and it is
expected to give better results when compared to experimental
measurements in intermediate $r_s$ values and even large momentum
limit where the Random Phase approximation calculation is no longer
accurate. Moreover, the small $k$ behavior of the charge excitation
spectrum is shown be equivalent with the bosonization findings.
Improvements of the theory will be necessary for a quantitative
study of the physical quantities and the correlation energy. As we
have mentioned in the main text, we only considered the low-order
elementary diagrams and the three-body corrections in the theory is
based on Fermi hypernetted-chain approximation. We believe inclusion
of the contributions of the fifth-order and higher elementary
diagrams will improve our results. Moreover as we mentioned in the
main text, we used a practical recipe of FHNC approach~\cite{lantto}
by replacing Jastrow product factor to the square of the Slater
determinant describing the non-interacting fermions gas
wave-functions. This recipe of FHNC reduces correlation respect to
the full FHNC-EL theory. Although it is little affect in high
dimensions but it may be important in the specific case of a 1D EL.

\begin{acknowledgments}
I am indebted to G. Senatore and M. Casula for providing me with
their DMC data reported in the figures. I also would like to thank
N. Nafari for useful discussions.
\end{acknowledgments}

\newpage
\begin{figure}
\begin{center}
\includegraphics[scale=0.6]{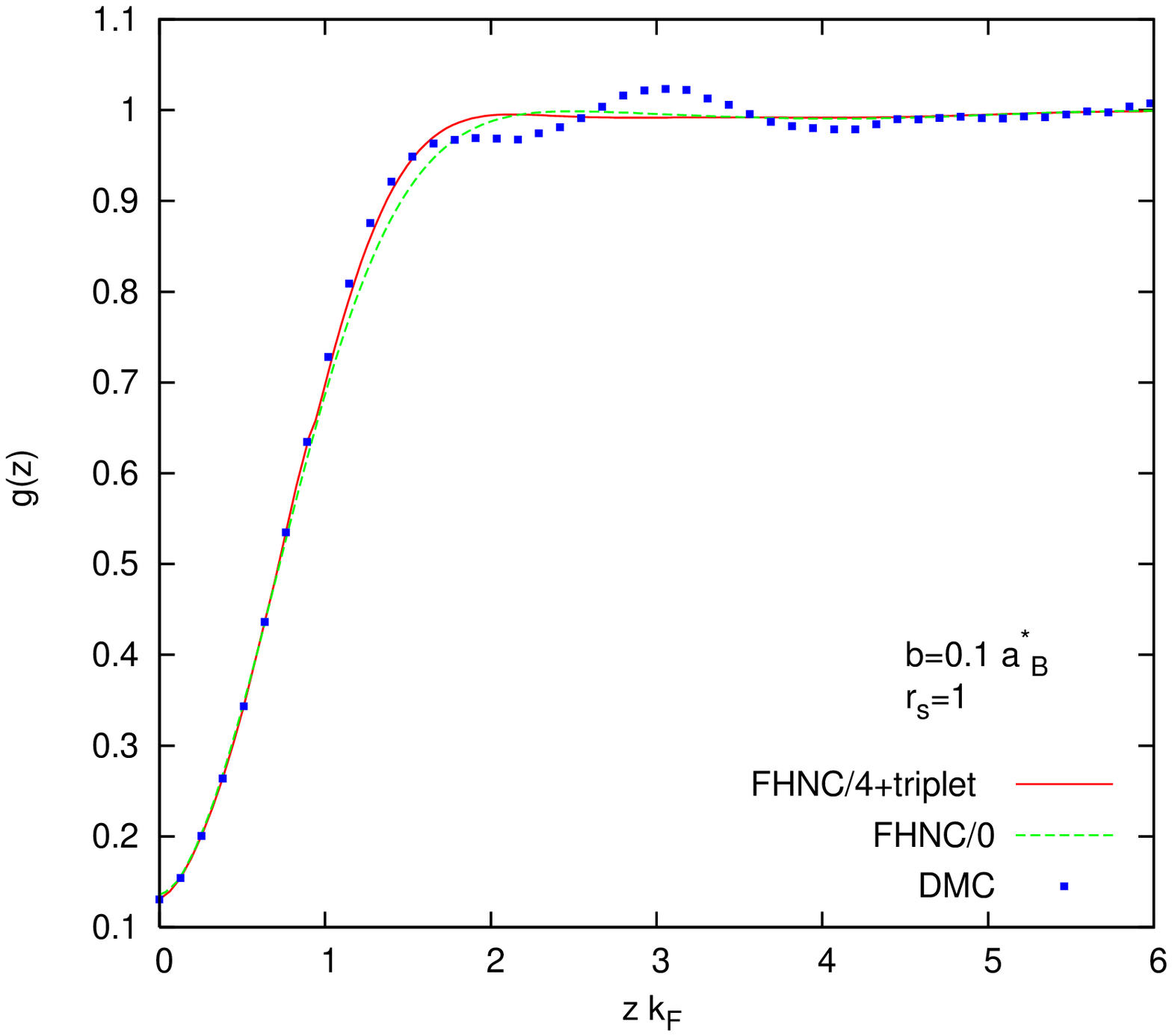}
\includegraphics[scale=0.6]{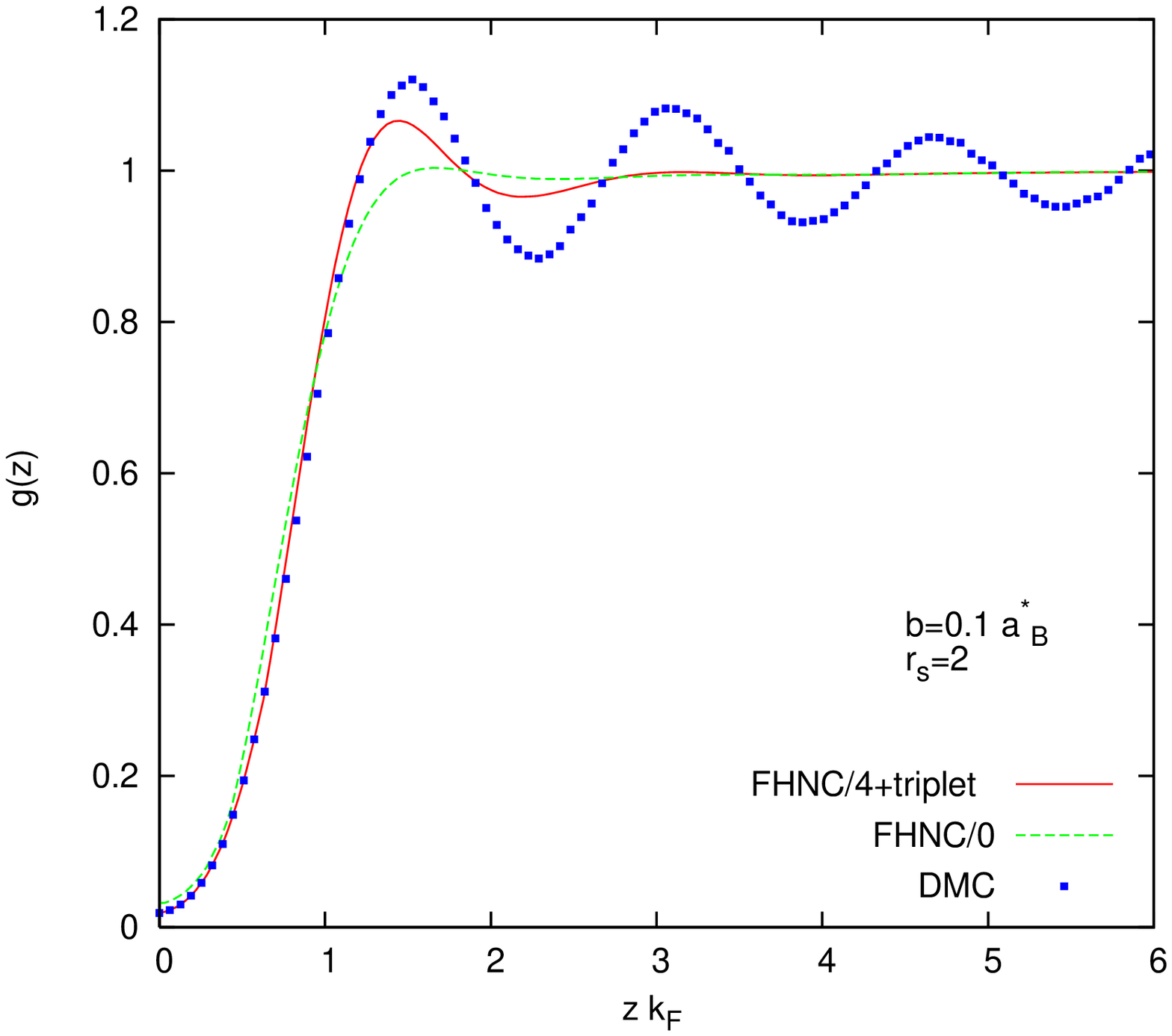}
\caption{(Color online)  The pair distribution function
$g(z)$ as a function of $z k_F $ for various $r_s=1$, ( top panel) and
$r_s=2$ (bottom panel) at $b=0.1 a^*_B$ comparing the FHNC/0 and FHNC$/4+$triplet
 approximations with DMC data of Casula {\it et al}~\cite{casula}. }
\end{center}
\end{figure}

\newpage
\begin{figure}
\begin{center}
\includegraphics[scale=0.6]{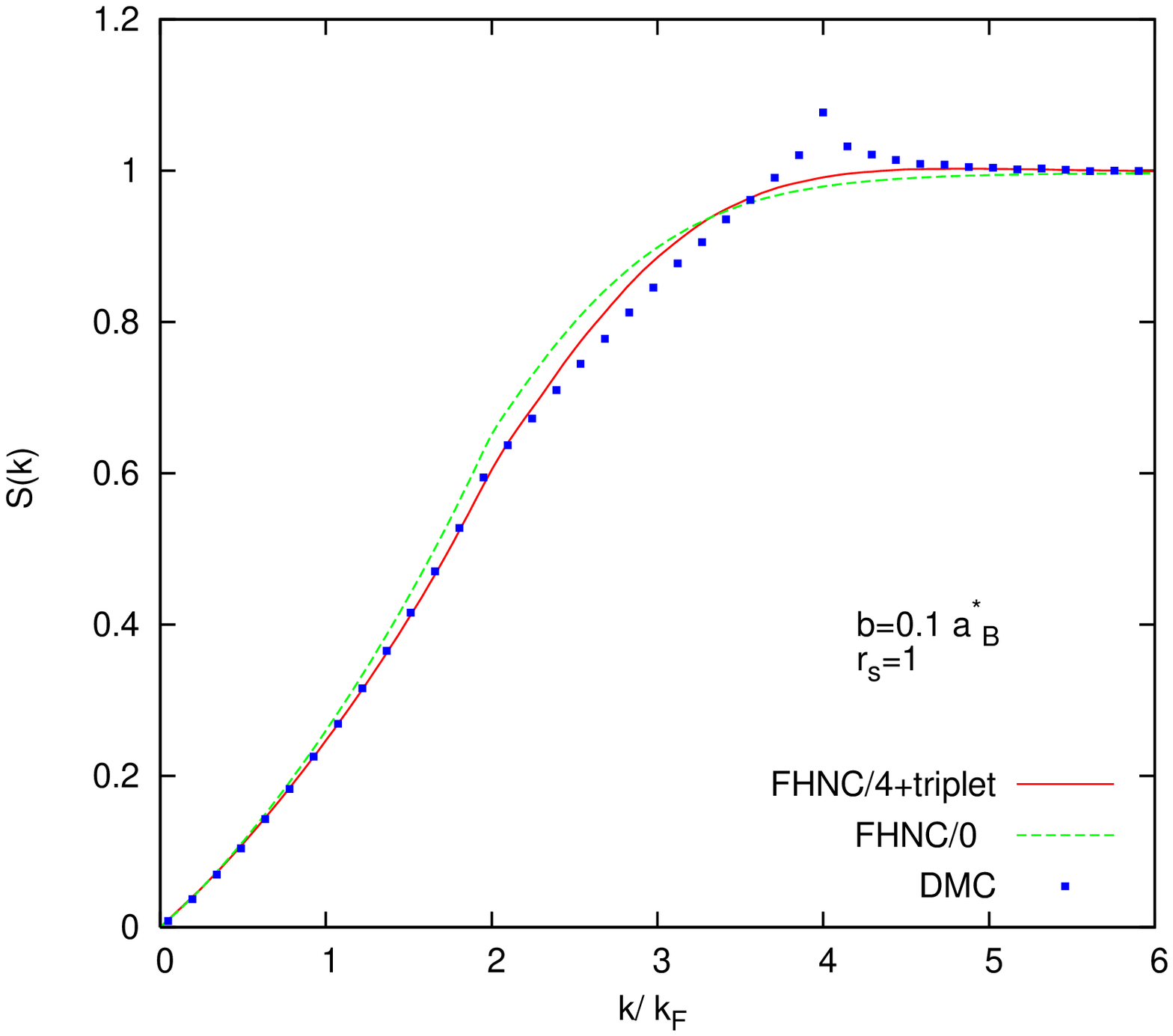}
\includegraphics[scale=0.6]{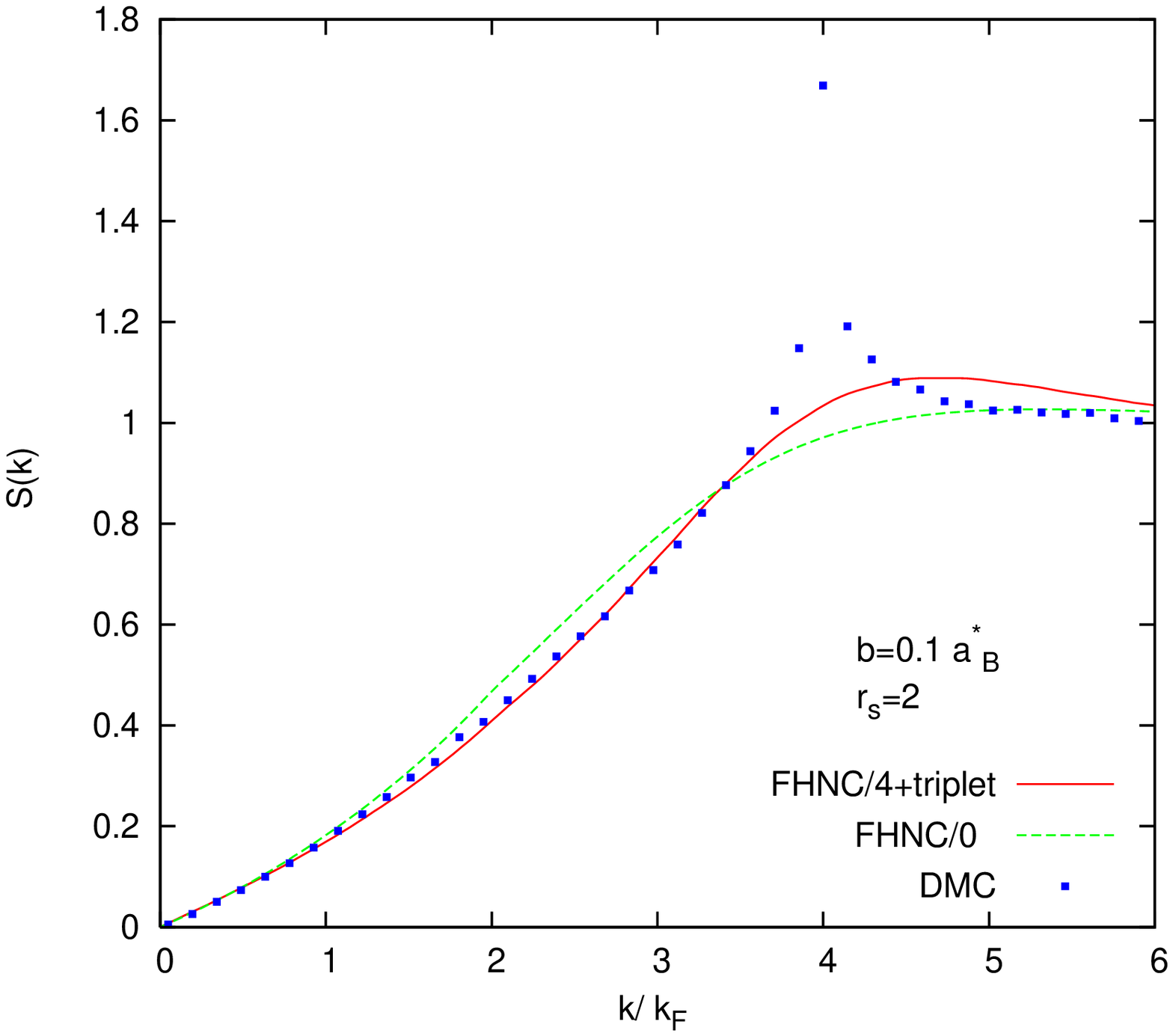}
\caption{(Color online)  The static structure function $S(k)$ as a
function of $k/ k_F $ for $r_s=1$ ( top panel), $r_s=2$ ( bottom
panel) and $b=0.1 a^*_B$ comparing both FHNC/0 and FHNC$/4+$triplet
approximations with DMC data of Casula {\it et al}~\cite{casula}. }
\end{center}
\end{figure}

\newpage
\begin{figure}
\begin{center}
\includegraphics[scale=0.6]{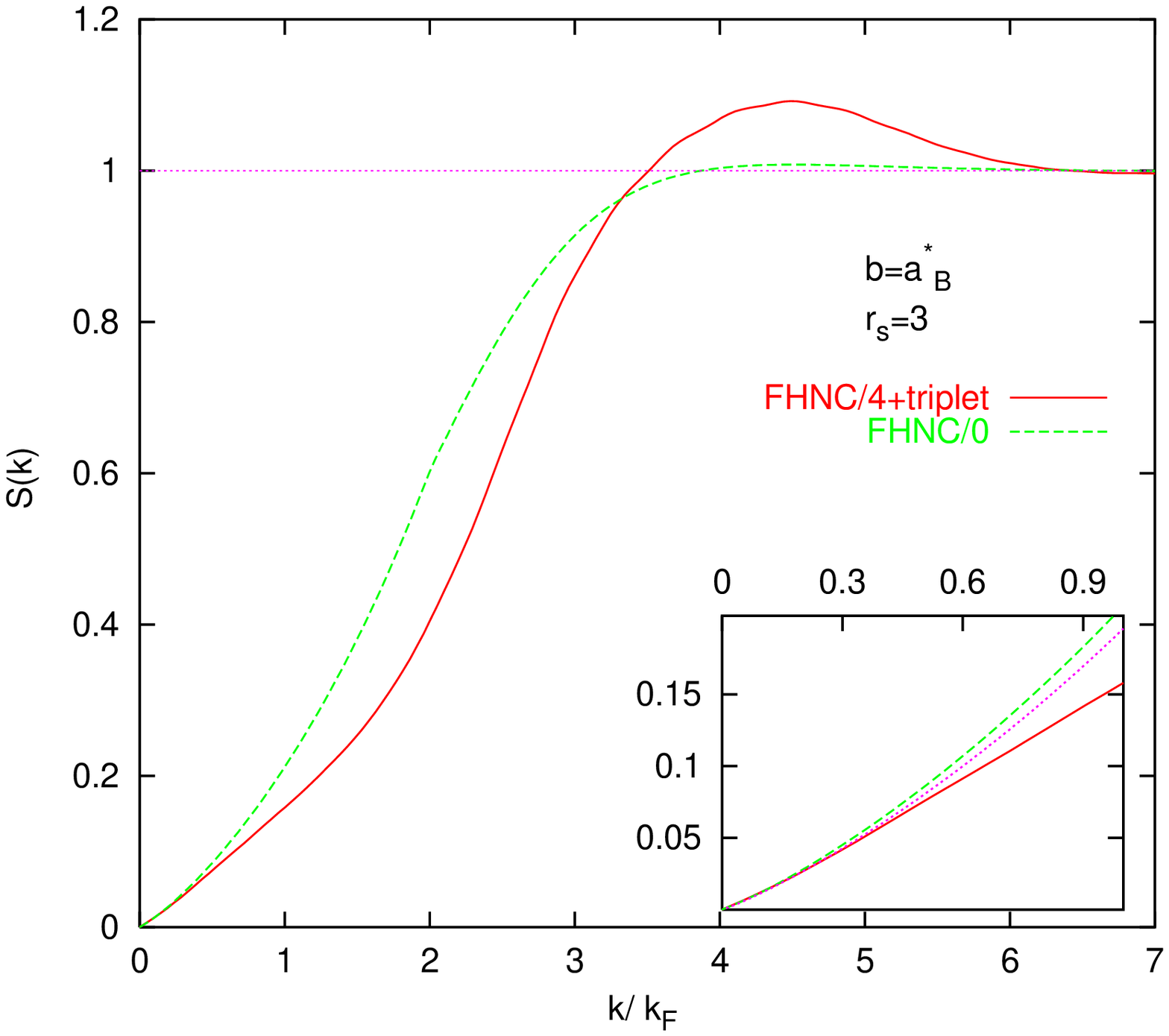}
\includegraphics[scale=0.6]{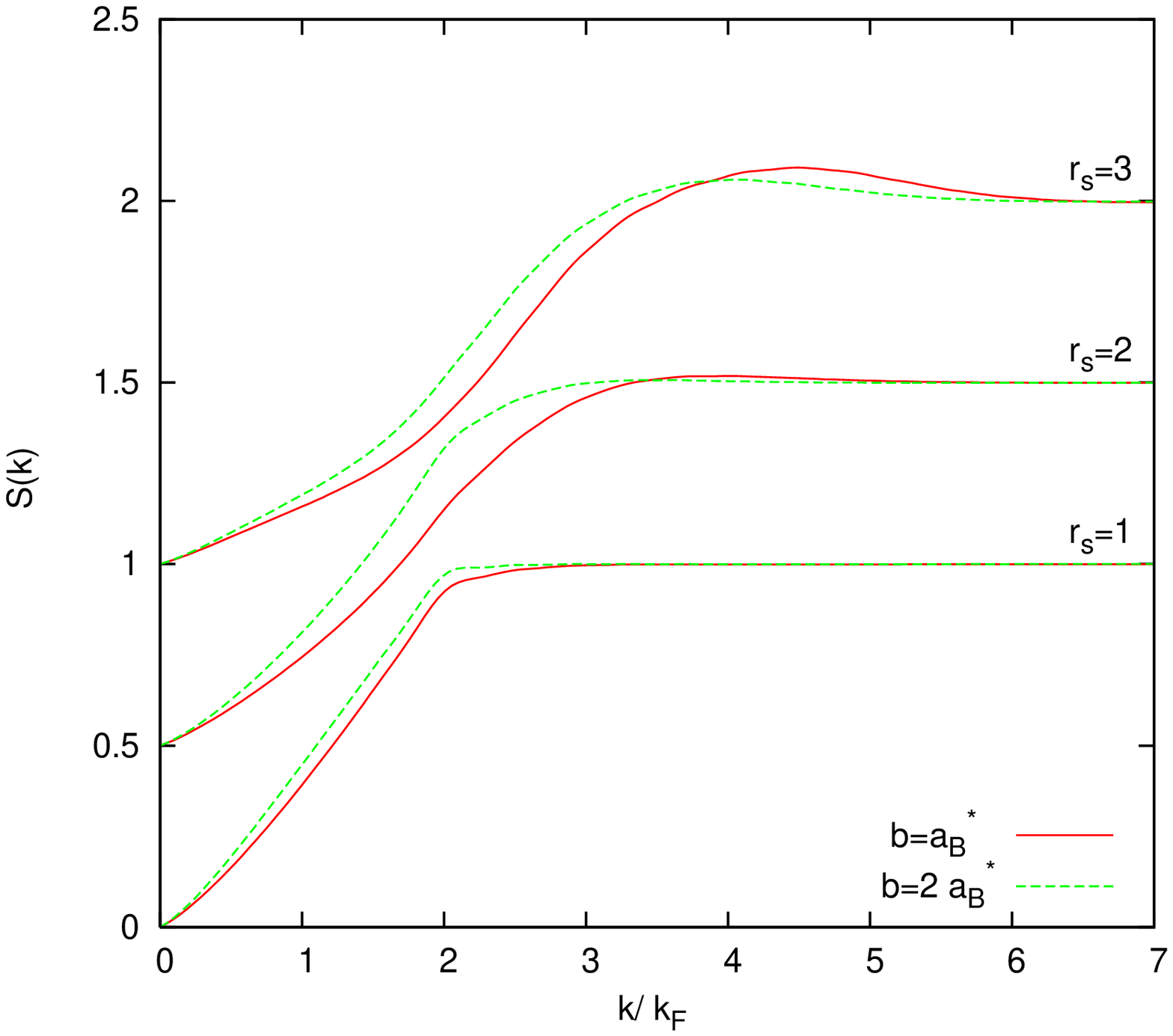}
\caption{(Color online) Top: The static structure function $S(k)$ as
a function of $k/ k_F $ for $r_s=3$ at $b=a^*_B$ calculating within
both FHNC/0 and FHNC/4$+$triplet approximations. In the inset the
long-wavelength behavior of $S(k)$ is shown in comparison with sum
rule relation (dotted curve). Bottom: The static structure function
$S(k)$ as a function of $k/ k_F $ for different $r_s$ and $b$ values
calculating within FHNC$/4+$triplet approximation. The curves at
$r_s=2$ and 3 have been shifted upwards by 0.5 and 1.0,
respectively.}
\end{center}
\end{figure}

\newpage
\begin{figure}
\begin{center}
\includegraphics[scale=0.6]{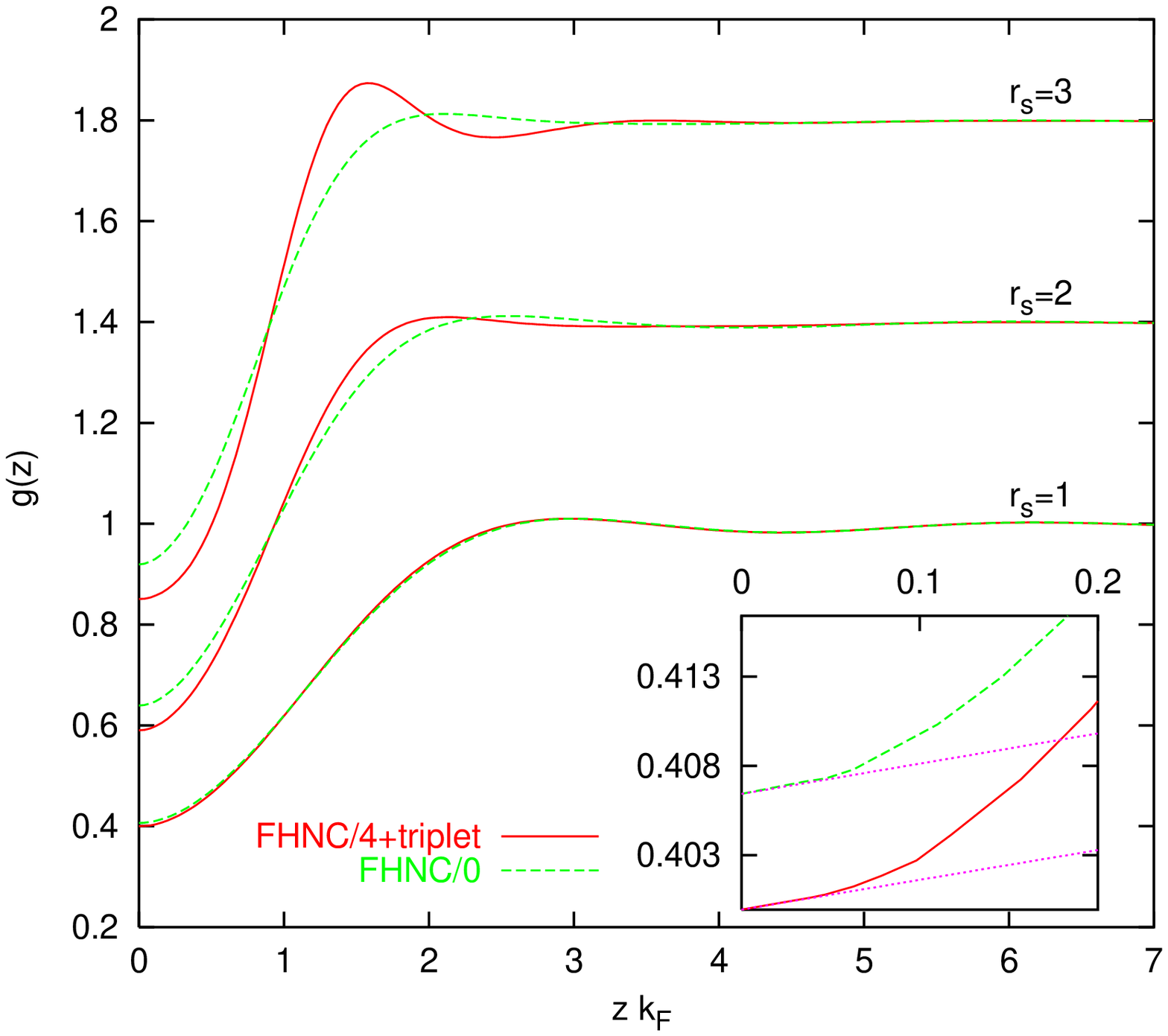}
\includegraphics[scale=0.6]{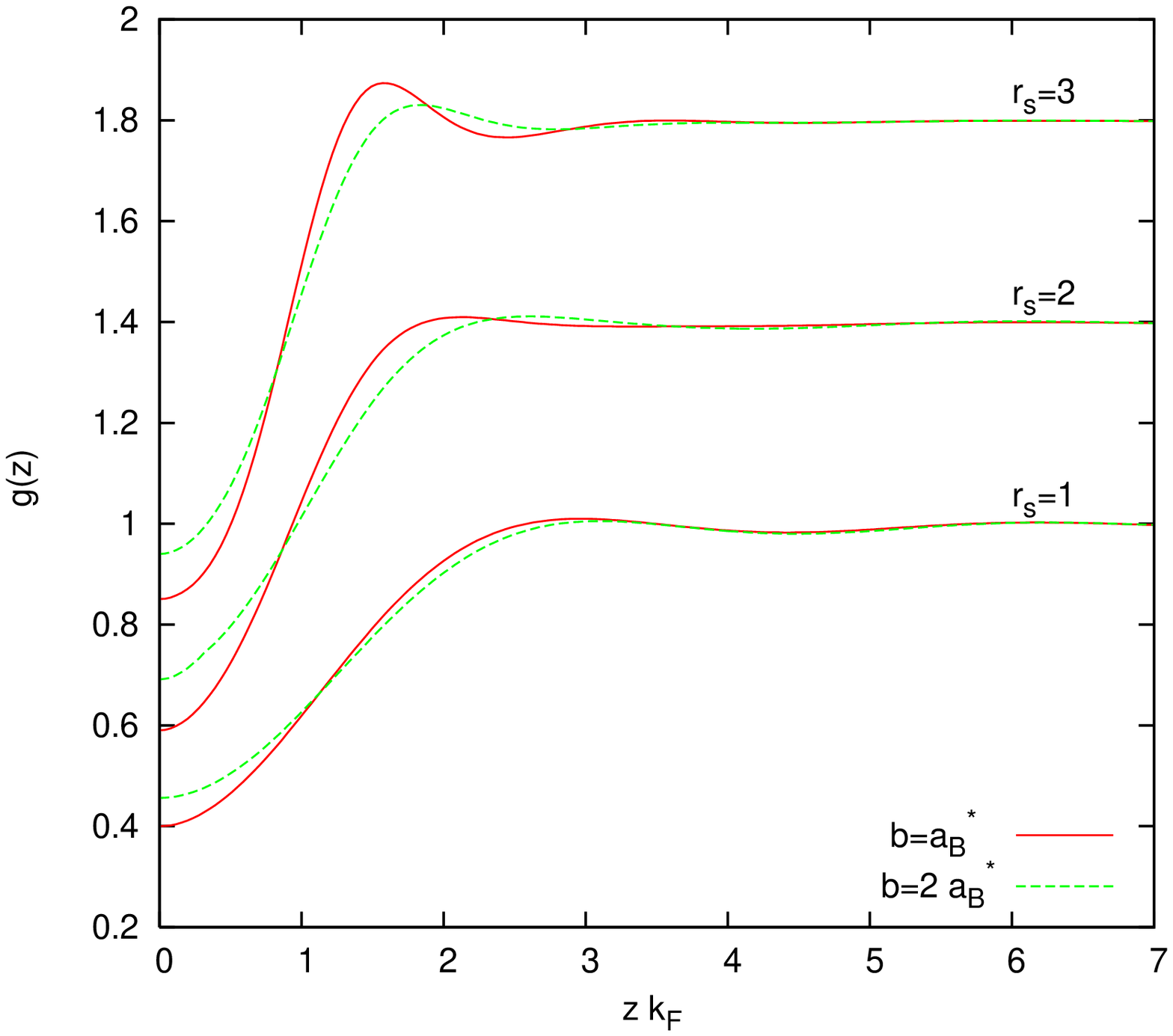}
\caption{(Color online) Top: The pair distribution function $g(z)$
as a function of $z k_F $ for various $r_s$ values at $b=a^*_B$
calculating within both FHNC/0 and FHNC$/4+$triplet approximations.
In the inset the cusp condition is shown (dotted curve). Bottom: The
pair distribution function $g(z)$ as a function of $z k_F $ for
different $r_s$ and $b$ values calculating within FHNC$/4+$triplet
approximation. The curves at $r_s=2$ and 3 have been shifted upwards
by 0.4 and 0.8, respectively.}
\end{center}
\end{figure}

\newpage
\begin{figure}
\begin{center}
\includegraphics[scale=0.6]{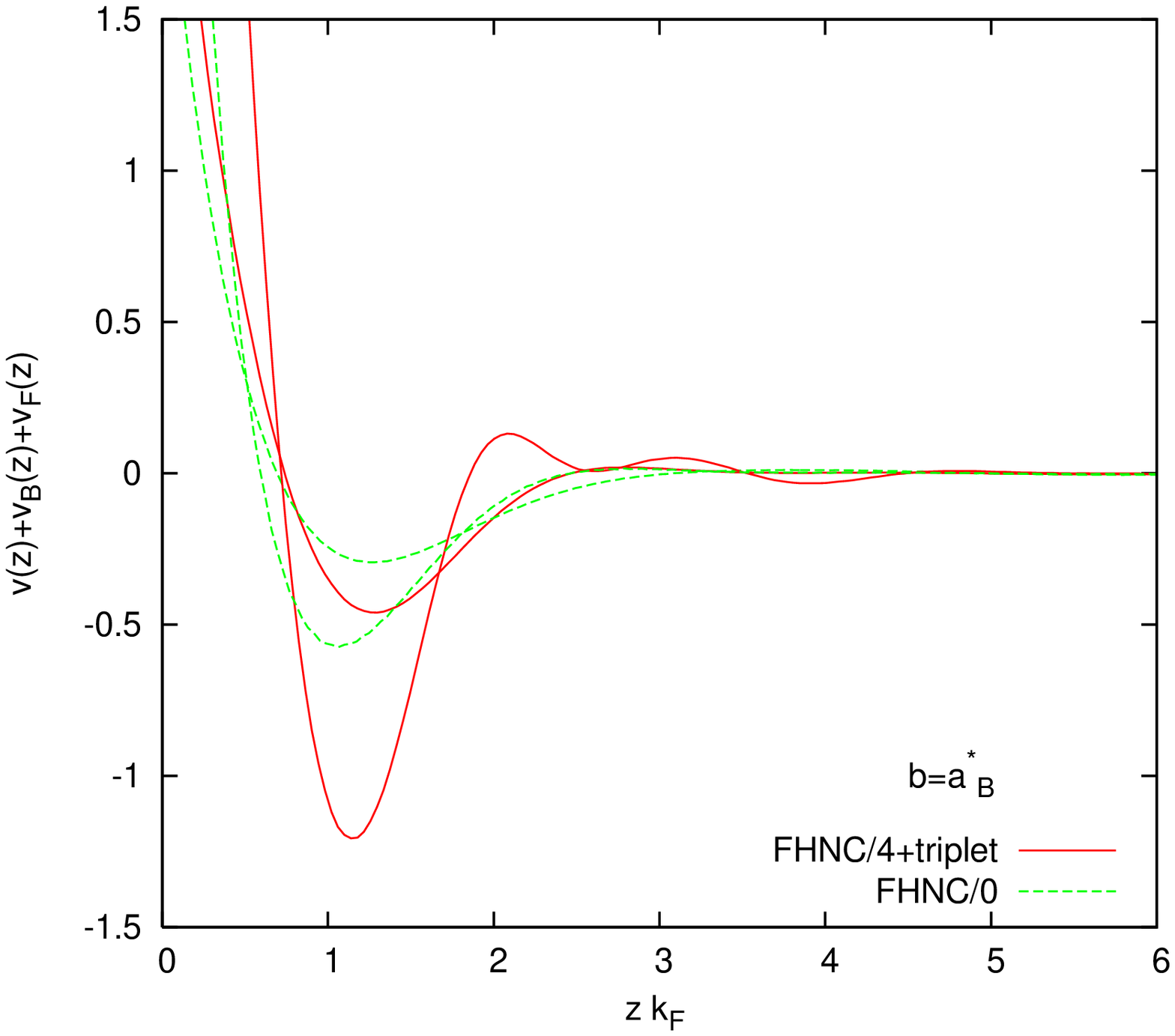}
\includegraphics[scale=0.6]{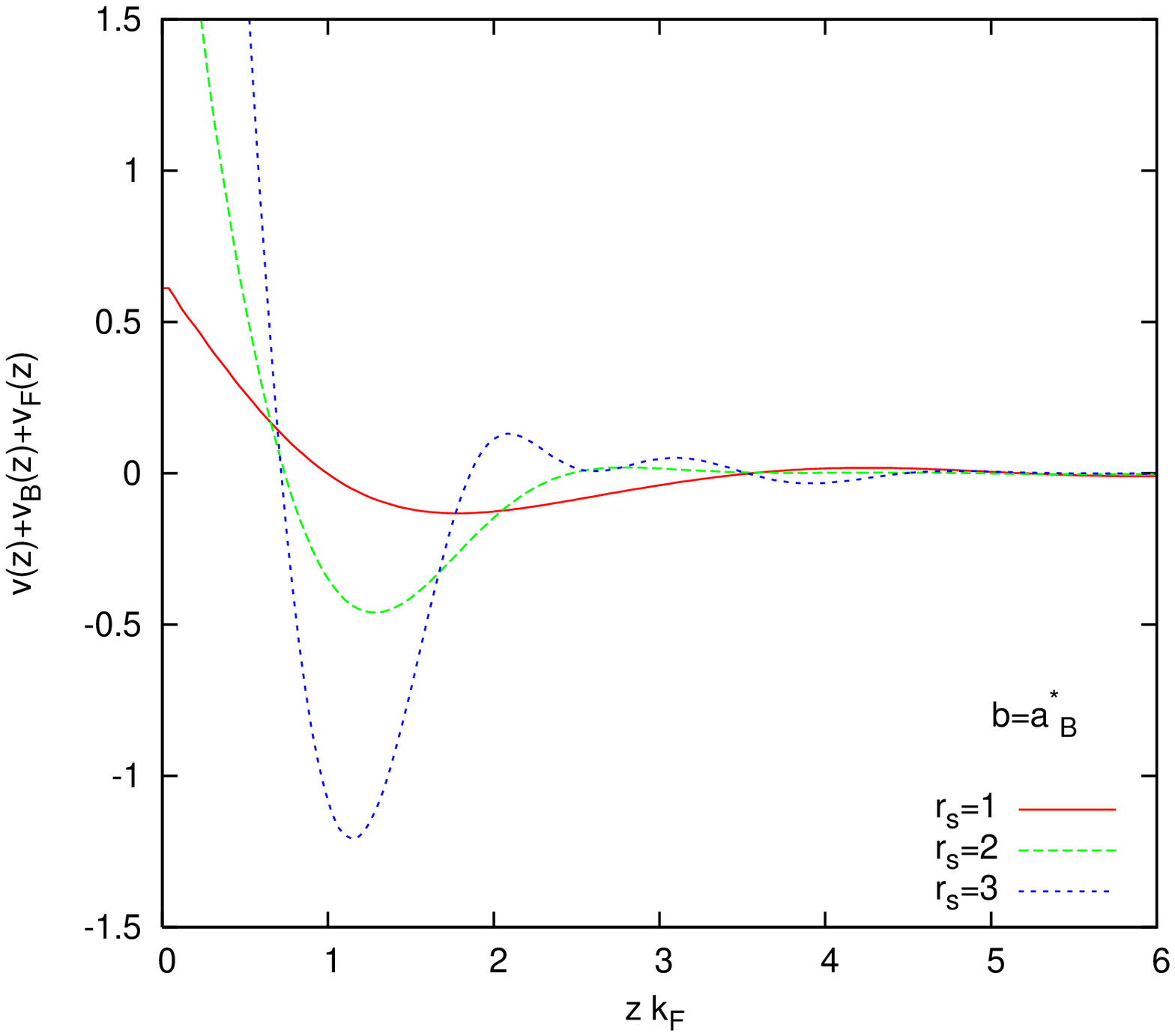}
\caption{(Color online) Top: The total potential
$v(z)+v_B(z)+v_F(z)$ as a function of $z k_F $ for $r_s=2$ and $3$ values at $b=a^*_B$ calculating within both FHNC/0 and FHNC$/4+$triplet approximations. Note that deeper potential refer
to large $r_s$ value.
Bottom: The total potential
$v(z)+v_B(z)+v_F(z)$ as a function of $z k_F $ for various $r_s$ at $b=a^*_B$  calculating within
FHNC$/4+$triplet approximation. Deeper potential indicate larger $r_s$ value.}
\end{center}
\end{figure}

\newpage
\begin{figure}
\begin{center}
\includegraphics[scale=0.6]{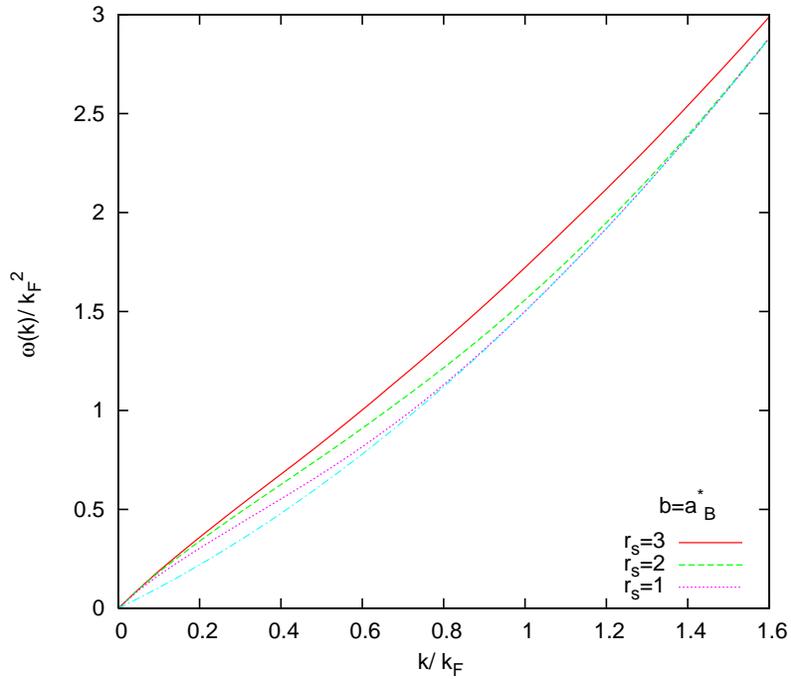}
\caption{(Color online) The charge excitation spectrume
$\omega(k)$ as a function of $k/ k_F $ for various $r_s$ values at $b=a^*_B$ obtaining from Eq.~(\ref{lff}) within the FHNC$/4+$triplet approximation. Dashed-dotted curve indicates the boundary of particle-hole continuum, $\beta_+$ .
}
\end{center}
\end{figure}

\newpage

\begin{table}
\caption{Correlation energy of the 1D EL in Ryd/electron. DMC from
Casula,~\cite{casula} STLS from Calmels and Gold~\cite{gold}}.
\begin{tabular}{lllllll}
\hline\hline $r_s$ &Various
calculations&$b=a^*_B$&$b=2a^*_B$\\
\hline
0.1         &DMC                              & -0.000463&-0.000110\\
            &Present work                    &-0.000459&-0.000107\\
            &STLS                                  &-0.000457&-0.000117\\ \hline
0.2          &DMC                              &-0.0016996&-0.000418\\
            &Present work                    &-0.001665&-0.000411\\
            &STLS                                  &-0.001645&-0.000431\\ \hline
0.4          &DMC                              &-0.00579&-0.001514\\
            &Present work                    &-0.00564&-0.001502\\
            &STLS                                  &-0.005449&-0.001492\\ \hline
0.6          &DMC                              &-0.01122&-0.003089\\
            &Present work                    &-0.01099&-0.002983\\
            &STLS                                  &-0.01044&-0.002955\\ \hline
0.8         &DMC                             &-0.01738&-0.00498\\
            &Present work                  &-0.01678&-0.00476\\
            &STLS                              &-0.01608&-0.00469\\ \hline
1.0         &DMC                               &-0.02394&-0.00709\\
            &Present work                  &-0.02296&-0.00687\\
            &STLS                              &-0.02202&-0.00662\\ \hline
2.0         &DMC                             &-0.05840&-0.01912\\
            &Present work                  &-0.05311&-0.01806\\
            &STLS                              &-0.04968&-0.01735\\ \hline
3.0         &DMC                             &-0.0856&-0.0322\\
            &Present work                  &-0.07330&-0.02952\\
            &STLS                              &-0.06862&-0.02744\\ \hline
4.0         &DMC                             &-0.09986&-0.04372\\
            &Present work                       &-0.08518&-0.03814\\
            &STLS                                  &-0.07978&-0.03556\\ \hline \hline
\end{tabular}
\end{table}

\end{document}